\documentstyle{article}

\Roman{section}

\input tcilatex

\QQQ{Language}{
American English
}

\begin{document}

\begin{titlepage}
\centerline{\large \bf New Types of Accessible Environmental Influence
on Neutrino Oscillation} 

\vspace{2.5cm}

\centerline{Chao-Hsi Chang$^{1,2}$, Hui-Shi Dong$^{3}$, Xi-Chen Feng$^{3}$,}
\vspace{0.3cm}
\centerline{Xue-Qian Li$^{1,3}$, Feng-Cai Ma$^{4}$ and Zhi-Jian Tao$^{5}$
}
\vspace{1cm}
{\small
\begin{center}

1.   CCAST  (World Laboratory) P.O.Box 8730 Beijing 100080, China.
\vspace{8pt}

2. Institute of Theoretical Physics, Academia Sinica, Beijing, 100080,
China.
\vspace{8pt}

3. Department of Physics, Nankai University,
Tianjin, 300071, China.
\vspace{8pt}

4. Department of Physics, Liaoning University,  Shenyang 110036, China.
\vspace{8pt}

5. Institute of High Energy Physics, Academia Sinica,
Beijing 100039, China.

\end{center}
}

\vspace{1cm}

\begin{center}
\begin{minipage}{12cm}
{\bf Abstract}
\vspace{1cm}

To describe possible effects of medium, 
besides the hamiltonian introduced by Mikheev, Smirnov
and Wolfenstein (MSW), in principle, there may exist new 
terms in the evolution equation for the neutrino oscillation
due to the environment.
Considering a two-energy-level
quantum system: $\nu_1-\nu_2$, two-species neutrino system
which is embedded in a matter environment,
we solve the neutrino evolution equation precisely with
the possible extra terms, especially, those
cause alternations between pure and mixed states of
the neutrino quantum system. We obtain some remarkable features for
the neutrino oscillations, which are
different from that of MSW and accessible to be observed 
in solar neutrino observation and/or in the planned 
long-baseline neutrino experiments.

\end{minipage}
\end{center}

\vspace{4mm}

{\bf PACS:} 14.60Pq, 03.65Bz, 05.75.-b

\end{titlepage}

\baselineskip 22pt \vspace{0.3cm}

To solve the solar neutrino deficit problem \cite{Smir}, there have been
many models which all assume neutrinos to be massive \cite{Lan}. And it has
been realized that the solar medium influence on the neutrino oscillation
may be crucial to solving the problem. Due to the influence of the solar
medium, the behavior of oscillation can be drastically modified and there
can be a resonance in matter if $\sin \theta _m\sim 1$, even though the
mixing angle in vacuum $\theta _v$ is very small. This is the well-known
Mikheev-Smirnov-Wolfenstein (MSW) mechanism \cite{Wol,Mik}. Besides the
solar matter effect, the Earth matter effect may also play a certain role in
the neutrino oscillation problem, i.e. the recently reported atmospheric
neutrino puzzle\cite{Hir} could be attributed to the medium effects either.

The key point of the oscillation scenario may be illustrated by a two-level
quantum system ($\nu_e,\nu_{\mu}$) or ($\nu_1, \nu_2$). The former are
eigenstates of weak-interaction and the later are that of mass matrix
respectively. In the scenario, the relevant quantities derived from quantum
field theory are transformed into a Hamiltonian of quantum mechanics (QM).
The species of neutrino produced by nuclear reactions is $\nu_e$ which can
be decomposed with respect to the basis of $\nu_1$ and $\nu_2$. Thus,
without the medium effects, its evolution is 
\begin{equation}
|\nu_e(t)>=\cos\theta_ve^{-iE_1t}|\nu_1>+\sin\theta_ve^{-iE_2t}|\nu_2>, 
\end{equation}
where $\theta_v$ is the mixing angle in vacuum and $E_1$, $E_2$ are energies
of $\nu_1$ and $\nu_2$. As neutrinos are light, we have $E_i\approx
k+m_i^2/2k\;\; (i=1,2)$, $k=|{\bf k}|$ and ${\bf k}$ is the three-momentum
of the neutrinos.

When MSW medium-effects are taken into account, the Hamiltonian should be
modified from $H$ to $H^{\prime}$ as 
\begin{equation}
\label{MSW}H=k+\left( 
\begin{array}{cc}
{\frac{m_1^2}{2k}} & 0 \\ 
0 & {\frac{m_2^2}{2k}} 
\end{array}
\right) \Longrightarrow H^{\prime }=k+\left( 
\begin{array}{cc}
{\frac{m_1^2}{2k}}-GN_e\cos {}^2\theta _v & -GN_e\sin \theta _v\cos \theta
_v \\ 
-GN_e\sin \theta _v\cos \theta _v & {\frac{m_2^2}{2k}}-GN_e\sin {}^2\theta_v 
\end{array}
\right) 
\end{equation}
where $GN_e$ is related to the electron density in a uniform matter
environment.

By solving the Schr\"{o}dinger equation for the neutrino evolution 
\begin{equation}
\label{old} i{\frac{d}{dt}}\left( 
\begin{array}{c}
\nu_1^{\prime} \\ 
\nu_2^{\prime} 
\end{array}
\right)=H^{\prime}\left( 
\begin{array}{c}
\nu_1^{\prime} \\ 
\nu_2^{\prime} 
\end{array}
\right) 
\end{equation}
one can see that a resonant structure of the mixing angle in medium $%
\theta_m $ is resulted in and the oscillation form becomes \cite{Wol,Mik} 
\begin{equation}
\label{MSW1} |<\nu_{\mu}|\nu_e(t)>|^2={\frac{1}{2}}\sin^2(2\theta_v)({\frac{%
l_m}{l_v}})^2 [1-\cos{\frac{2\pi t}{l_m}}], 
\end{equation}
where 
\begin{equation}
l_m=l_v(k)[1+({\frac{l_v(k)}{l_0}})^2-2\cos(2\theta_v)({\frac{l_v(k)}{l_0}}%
)]^{-1/2}, 
\end{equation}
and 
$$
l_0={\frac{2\pi}{GN_e}},\;\;\; l_v={\frac{4\pi k}{m_1^2-m_2^2}}.%
$$
The mixing angle and period of the neutrino oscillation are modified
accordingly.

Recently, some authors have studied various extensions of quantum mechanics
(QM) (or sometimes called violation of QM), such as due to existence of
`micro black holes', a non-hermitian Hamiltonian is induced, thus a quantum
system of pure state would turn into a mixed state, and CP, even CPT
violation may be resulted\cite{Ellis,Huet}. From a different motivation,
Reznik studied a quantum system submerged in a `proper surrounding
environment' and he found that instead of a non-reversible evolution of the
quantum system from a pure state to a mixed, there may be a periodic
transition between pure and mixed states \cite{Reznik}.

In fact, neutrino propagating in the dense medium of the Sun may be such a
system. The oscillation between two species of the neutrinos, in principle,
may be affected by the environment on its way to the surface of the Sun.
First of all, we would like to consider the `extension' by Reznik, and
following him we write down a generalized Hamiltonian and find the solution
of the evolution equation. One can note that it is indeed an extension of
the Mikheev-Smirnov-Wolfenstein's result.

In the following, we briefly recall Reznik's work and present our derivation
of a modified expression for $|<\nu_{\mu}|\nu_e(t)>|^2$, as well as
demonstrate how the modification practically affects the oscillation rate.\\

\noindent{\it Formulation} \hskip 0.5cm (i) Unitary evolution between pure
and mixed states for a quantum system embedded in a matter environment.

The quantum system can be described by a density matrix $\rho $ in general.
If the system is in a pure state, one has $Tr\rho^2=Tr\rho=1$, instead, when
it is in a mixed state, $Tr \rho^2<1$. In quantum mechanics, the
Schr\"{o}dinger equation is 
\begin{equation}
\label{Sch} i{\frac{d}{dt}}\rho=[H,\rho]. 
\end{equation}

In order to consider the environmental influence due to matter in very
general cases, especially a possible unitary evolution of the system between
pure and mixed states, Reznik proposed an operator $\hat{\rho}$ which
relates to $\rho$, the reduced density matrix of the system, as 
\begin{equation}
\rho\equiv\hat{\rho}\hat{\rho}^{\dagger}, 
\end{equation}
whereas $\hat\rho^{\dagger}\hat\rho$ (in general $\neq \rho$) stands for the
density matrix of the environment. At absence of an extra QM extension term, 
$\hat\rho$ still satisfies the same Liouville equation as (\ref{Sch}).

At the time $t=0$, the system is supposed to stay in a pure state, so 
\begin{equation}
\rho(t=0)\equiv |\psi_0><\psi_0|=\hat\rho=\hat\rho^{\dagger}. 
\end{equation}

Reznik shows that the equation (\ref{Sch}) can be generalized to 
\begin{equation}
\label{hat} i{\frac{d}{dt}}\hat\rho=[H,\hat\rho]+L\hat\rho+\hat\rho
R+g_{ij}K_i\hat\rho K^{\prime}_j, 
\end{equation}
where $L,\; R,\;, K_i$ and $K^{\prime}_j$ all are hermitian operators, $%
g_{ij}$ are real parameters.

It is noted that as long as the $K_j$ in eq.(\ref{hat}) is not a unit matrix
($\sigma _0$), i.e. it is not a trivial case, there cannot be a simple
equation for the density matrix $\rho $. Thus people sometimes call the last
term in eq.(\ref{hat}) as an extension of QM, whereas being different in
nature, a corresponding extra term is named as the violation term of QM in
the case of ref.\cite{Ellis}. In the present case, rewriting eq.(\ref{hat}),
we have 
\begin{equation}
i{\frac d{dt}}\rho =[H+L,\rho ]+g_{ij}(K_i\hat \rho K_j^{\prime }\hat \rho
^{\dagger }-h.c), 
\end{equation}
and in the last part of the equation, $\rho $ does not appear as an
independent variable directly at all (see below for more details). We can
also easily prove that a resultant $\rho \equiv \hat \rho \hat \rho
^{\dagger }$ describes a mixed state as $Tr(\rho ^2)<1$ in the case. In
fact, if there is no the last extension term, $d/dt(Tr\rho ^2)=0$ always
holds, namely, the system remains in either pure or mixed state forever. It
never transits from a pure state to a mixed or vice versa. Whereas when the
`extra' term exists, it is easy to prove $d/dt(Tr\rho ^2)\neq 0$, and it
indicates that an alternation 
\footnote{Throughout the paper we adopt the ward
`alternation' here instead of `oscillation' 
adopted by \cite{Reznik}, to avoid possible
confution with the `oscillation' for neutrinos.} between pure and mixed
states occurs.

(ii) Let us consider a system, that only two species of neutrinos are
involved, thus they can be treated as a two-energy-level quantum system.

Following ref.\cite{Reznik} and considering the situation for neutrinos, we
may write down the term $H$ appearing in eq.(\ref{hat}) directly as 
\begin{equation}
H=k+\left( 
\begin{array}{cc}
{\frac{m_1^2}{2k}} & 0 \\ 
0 & {\frac{m_2^2}{2k}} 
\end{array}
\right) 
\end{equation}
in the basis of $|\nu_1>$ and $|\nu_2>$. The trivial part $k$ is a unit
matrix and can be omitted in our later calculations. After a transformation $%
\hat\rho\rightarrow\hat\rho\; U$ with 
$$
U=exp[-i\int^t(R-H)dt^{\prime}],%
$$
according to ref.\cite{Reznik} the eq.(\ref{hat}) is turned to 
\begin{equation}
\label{new} i{\frac{d}{dt}}\hat\rho=\tilde H\hat\rho+g_{ij}K_i\hat\rho\tilde
K^{\prime}_j 
\end{equation}
and 
$$
\tilde H=H+L.%
$$
The extra Hamiltonian L is hermitian which is induced by the environment,
therefore can be absorbed into the original Hamiltonian H.

Comparing eq.(\ref{old}) and eq.(\ref{new}), $\tilde H$ can be identified as 
$H^{\prime}$ and one can easily believe that the extra L is nothing new, but
the Hamiltonian derived by Mikheev, Smirnov and Wolfenstein. Thus 
\begin{equation}
L=\left( 
\begin{array}{cc}
-GN_e\cos^2\theta_v & -GN_e\sin\theta_v\cos\theta_v \\ 
-GN_e\sin\theta_v\cos\theta_v & -GN_e\sin^2\theta_v 
\end{array}
\right). 
\end{equation}
Therefore $\tilde H\equiv H+L$ is exactly the hamiltonian $H^{\prime}$ in
eq. (\ref{MSW}).

However, one can notice that there is one more extra term in eq.(\ref{new}),
i.e. $g_{ij}K_i\hat\rho \tilde K_j^{\prime}$ which violates the hermiticity
of $\hat\rho $ (but not $\rho$) and it would influence the behavior of the
neutrino oscillation. For simplicity, later on we will write this term just
as $K\hat\rho \tilde K^{\prime}$ and the free parameters are absorbed in $K$
and $\tilde K^{\prime}$.

(iii) $\tilde H=H+L$ is a $2\times 2$ hermitian matrix, so it can be
decomposed as 
\begin{equation}
\tilde H=a_0\sigma_0+a_1\sigma_1+a_2\sigma_2+a_3\sigma_3. 
\end{equation}
The first term in this expression is trivial, so we will neglect it in later
calculations, and furthermore from the explicit form of $\tilde H$ given in
eqs.(2) and (13), take $a_2=0$ in later calculations.

Due to energy conservation, $[K,\tilde H]=0$ must be satisfied, a general
form of $K$ should be 
$$
\alpha (a_1\sigma_1+a_3\sigma_3), 
$$
where $\alpha$ is an arbitrary parameter. $\tilde K^{\prime}$, being a
hermitian $2\times 2$ matrix, can be written as 
$$
\tilde K^{\prime}=\lambda_1\sigma_1+\lambda_2\sigma_2+\lambda_3\sigma_3, 
$$
where a possible but trivial term, $\lambda_0\sigma_0$, has been dropped out.

Just for illustration and convenience for later discussions, following ref.%
\cite{Reznik}, we choose the simplest form for $\tilde K^{\prime}$ as $%
\lambda\sigma_3$ (i.e. $\lambda_1=\lambda_2=0$ and $\lambda_3=\lambda$).
Thus the extension term becomes 
\begin{equation}
\label{change} K\hat\rho\tilde
K^{\prime}=(a_1\sigma_1+a_3\sigma_3)\hat\rho\lambda\sigma_3, 
\end{equation}
where the arbitrary parameter $\alpha$ is absorbed into $\lambda$. So far,
we cannot derive this term from a well-established theory yet, so that we
leave it aside as an open question.

As usually adopted method\cite{Ellis,Huet,Reznik}, we write $\hat\rho$ in a
four-vector form. In fact, this step is not trivial at all because of the QM
extension term, (below we will give more discussions on this point). One can
decompose $\hat\rho$ as 
$$
\hat\rho=\hat\rho_0\sigma_0+\hat\rho_1\sigma_1+
\hat\rho_2\sigma_2+\hat\rho_3\sigma_3.%
$$
Thus 
$$
{\frac{d}{dt}}\hat\rho=\tilde H\hat\rho+K\hat\rho\tilde K^{\prime}%
$$
would become a form 
\begin{equation}
\label{four} i{\frac{d}{dt}}\hat\rho_a=(\tilde H_{ab}^{qm}+\delta
H_{ab})\hat\rho_b, \;\;\;\; a,b=0,1,2,3, 
\end{equation}
where both $\tilde H^{qm}$ and $\delta H$ are $4\times 4$ matrices which
correspond to $\tilde H=H+L$ and $K\hat\rho\tilde K^{\prime}$ respectively.
Thus we have 
\begin{equation}
\label{eq} i{\frac{d}{dt}}\left( 
\begin{array}{c}
\hat\rho_0 \\ 
\hat\rho_1 \\ 
\hat\rho_2 \\ 
\hat\rho_3 
\end{array}
\right)=\left( 
\begin{array}{cccc}
a_0+\lambda a_3 & a_1 & i\lambda a_1 & a_3 \\ 
a_1 & a_0-\lambda a_3 & -ia_3 & \lambda a_1 \\ 
-i\lambda a_1 & ia_3 & a_0-\lambda a_3 & -ia_1 \\ 
a_3 & \lambda a_1 & ia_1 & a_0+\lambda a_3 
\end{array}
\right)\left( 
\begin{array}{c}
\hat\rho_0 \\ 
\hat\rho_1 \\ 
\hat\rho_2 \\ 
\hat\rho_3 
\end{array}
\right)\equiv H_{(4\times 4)}\hat\rho^V. 
\end{equation}
It is easy to check that in the $(4\times 4)$ form, $H_{(4\times 4)}=
(\tilde H_{ab}^{qm}+\delta H_{ab})_{4\times 4}$ still retains hermiticity.

(iv) Solution of eq.(\ref{eq}) can be obtained by diagonalizing the $4\times
4$ matrix. Because it is a hermitian matrix, so it can be diagonalized via a
unitary matrix $V$, as $diag (H_{(4\times 4)})=VH_{(4\times 4)} V^{\dagger}$%
. Thus 
\begin{equation}
\label{sol} \hat\rho_a(t)=\sum_{b,c}V^{\dagger}_{ab}V_{bc}\hat%
\rho_c(t=0)e^{-i\beta_bt}, 
\end{equation}
where $\beta_b$ are eigenvalues of $H_{(4\times 4)}$ in eq.(\ref{eq}). Since
at the moment t=0, the system is at a pure state, $\hat\rho(t=0)=\rho(t=0)$.
Without losing generality we assume a pure $\nu_e$ at the initial state,
thus 
\begin{equation}
\rho(t=0)=\hat\rho(t=0)=|\nu_e><\nu_e|=\left( 
\begin{array}{cc}
\cos^2\theta_v & -\sin\theta_v\cos\theta_v \\ 
-\sin\theta_v\cos\theta_v & \sin^2\theta_v 
\end{array}
\right). 
\end{equation}
Converting the initial condition in $2\times 2$ matrix form into a
four-vector form and substituting those $\hat\rho_c(t=0)$ into the solution (%
\ref{sol}), we obtain $\hat\rho_a(t)$ and $\rho_e(t)\equiv
\hat\rho_e(t)\hat\rho^{\dagger}_e(t)$ in $(2\times 2)$ matrix form.

In fact, a complete form of $K\hat\rho \tilde
K^{\prime}=\alpha(a_1\sigma_1+a_3
\sigma_3)\hat\rho(\lambda_1\sigma_1+\lambda_2\sigma_2+\lambda_3\sigma_3)$
which replaces the simple one in eq.(\ref{change}), does not change the
whole physical picture. Our full derivation shows that to obtain the
formulation with all the $\lambda_i$ (i=1,2,3) existing, one only needs to
re-write $\lambda$ of eq.(\ref{change}) in a complicated combination of $%
\lambda_i$ whereas the general form remains unchanged. So a solution derived
from the simple expression does not lose generality. Here we omit the
details to save space.

(v) The oscillation rate is defined as 
\begin{equation}
\label{rate} P_{\nu_e\rightarrow\nu_{\mu}}(t)\equiv
|<\nu_{\mu}|\nu_e(t)>|^2= Tr(\rho^{\dagger}_{\mu}\rho_e(t)). 
\end{equation}
Thus we have the final expression of $P$ as 
\begin{equation}
\label{res} P(l)={\frac{1}{2}}\sin^22\theta_v({\frac{l_m}{l_v}})^2[1-\cos({\ 
\frac{2\pi l}{l_m}}) \cos({\frac{2\pi\lambda l}{l_m}})+\cos 2\theta_v\sin({\ 
\frac{2\pi l}{l_m}}) \sin({\frac{2\pi\lambda l}{l_m}})], 
\end{equation}
where $l$ is the distance from the production site of $\nu_e$ to the
detection point. Note that our $\lambda$ is dimensionless and it is slightly
different from Reznik's notation where his $\lambda$ has a dimension of
energy. It is a modified expression of the neutrino oscillation in medium
and simply an extension from that given by Mikheev, Smirnov and Wolfenstein.

Note that we have checked that the extension of QM which we are considering,
is of unitarity. Because of conservation of possibilities the constraint 
$$
P_{\nu_e\rightarrow\nu_{\mu}}(t) + P_{\nu_e\rightarrow\nu_e}(t)=1%
$$
is satisfied exactly.

As $\lambda=0$, which implies the extra term $K\hat\rho K^{\prime}$
disappears, the expression reduces back to the form given in ref.\cite{Wol}
and \cite{Mik}. For non-zero $\lambda$ values, the oscillation is not simply
harmonic, but in a modified mode. In general the oscillation period is
shortened, so for the solar neutrino case the result favors a larger
reduction rate of the electron neutrinos produced in the Sun.

(vi) To compare various possibilities, let us consider another type of
environmental influence on the neutrino oscillation due to the `micro black
holes everywhere' although the authors of ref.\cite{Ellis} originally
proposed it for the quantum system $K^0\to \bar{K^0}$ mixing. According to
the mechanism given in ref.\cite{Ellis}, the oscillation rate may be
obtained: 
\begin{equation}
\label{black}P(t)=|<\nu _\mu |\nu _e(t)>|^2={\frac 12}\sin {}^22\theta
_v(1-e^{{\frac{-(\alpha +\gamma )}2}t}\cos {\frac{2\pi t}{l_v}}), 
\end{equation}
where an approximation 
$$
{\frac{\Delta}{2k}}\gg |\alpha|,\; |\beta|,\; |\gamma| \;\;\;\;\;
(or\;\alpha\sim\beta\gg |\beta|)%
$$
is required.

That is a modified vacuum oscillation and the damping factor indicates an
energy loss at the evolution process. It is understood that as the neutrinos
propagate in the physical vacuum full with micro black holes, their
evolution behavior are affected by the black holes. Simultaneously the
quantum system turns from a pure state into a mixed state and this trend is
non-reversible since the black holes never release anything out (here we do
not refer to the quantum tunneling effect of black holes). The new behaviors
of the system evolution will be discussed in our coming work \cite{Chang}.\\

\noindent{\it Discussions and implications} \hspace{0.5cm} (i) It is
definite that the solar environment can significantly influence the neutrino
oscillation. Mikheev, Smirnov and Wolfenstein's work accounted such effect
by adding an extra hamiltonian to the system. Here in a more general
framework, as Reznik suggested, there is a natural way to include the
influence from environment, especially the system may be oscillating between
pure and mixed states. Namely, the environment interacts with the quantum
system in an unknown and complicated way. Such an interaction is not only
depicted by a hermitian Hamiltonian, but also by an additional term, which
violates the hermiticity of $\hat\rho$. It makes the system evolve between
pure and mixed states. The MSW's Hamiltonian only results in a hermitian
part of $\hat\rho$. It is also noted that the interaction of the neutrino
system with the magnetic field of the Sun \cite{Vol} can also induces an
external environmental effect but it may also be absorbed into $L$. Only the
extension part can cause the system to alternate between pure and mixed
states\cite{Reznik}.

Here the situation is different from that of the `micro black holes'. There
are interactions between the neutrino system and the environment, although
we cannot theoretically derive them precisely at present, In general the
energy of the whole system is conserved, even though it evolves from a pure
state into a mixed state and vise versa. From the analytic expression (\ref
{res}), we also observe that the oscillation behavior of the neutrino system
with the QM extension term is quite different from that without it. In the
regular QM region where $K\hat\rho K^{\prime}$ is absent (i.e. $\lambda=0$),
the oscillation rate $|<\nu_{\mu}|\nu_e(t) >|^2$ can vary between 0 and $%
\sin^2(2\theta_m)$ (or $\sin^2(2\theta_v)$ for the vacuum case) \cite
{Wol,Mik}, on the contrary, due to the presence of the QM extension term, $%
|<\nu_{\mu}|\nu_e(t)>|^2$ can never be zero except at the moment t=0.

For the solar neutrino problem since the initial $\nu _e$ is produced all
over the Sun and the distribution of the chemical elements in the Sun
varies, one must integrate the transition probability over the whole
production region and energy spectrum of neutrinos. It is 
$$
{\frac 16}\int_E\int_{-L}^{+L}P(L-l)\eta (L-l,E)dldE, 
$$
where the origin is chosen at the center of the Sun, and the time, which the
neutrino takes for traveling from $l$ to the surface of the Sun, is $t=l/v
\sim l/c$ and $v$ is the speed of the neutrino, which is close to that of
light $c$, as long as neutrino masses are supposed to be very tiny. The
factor 1/6 accounts for an average probability in the propagation direction
to the Earth. If the distribution of the neutrino production $\eta (L-l,E)$
does not vary drastically, it can be approximately treated as a constant,
the integration over sites can be carried out analytically.

To solve the solar neutrino problem with MSW solution as suggested by many
literatures, $\Delta\equiv m_1^2-m_2^2$ is comparatively large. We have $%
L\gg l_m$\cite{Smir1}, i.e. the solar size is much longer than the
oscillation length. In the case the extra contribution in eq.(21) would
disappear as well as the oscillation term in eq.(4) which is proportional to 
${\frac{2l_m}{2\pi L}}\cos(2\pi L/l_m)$, unless $\lambda$ is close to unity.
If $\lambda\sim 1$, the modified expression (21) can be averaged as 
\begin{equation}
{\frac{1}{L}}\int_{-L}^{+L}P(L-l)dl={\frac{1}{2}}\sin^2\theta_v({\frac{l_m}{%
l_v}})^2 \cos^2\theta_v. 
\end{equation}
Note that, as long as $\lambda\neq 1$ and $l_m\ll L$, the extension term $%
K\hat\rho\tilde K^{\prime}$ does not change the physics obtained in the
regular framework without such a term, whereas $\lambda\rightarrow 1$, an
extra factor $\cos^2\theta_v$ would be attached to the original expression
that is observable. If $\theta_v$ is small, this effect can also be
neglected.

In fact, the medium is not uniform, i.e. $\eta (L-l,E)$ is not a constant
over $l$ and $\lambda $ may be environment dependent, there may be some
effects which can influence the data fitting of $\theta _m$, but to take
into accounts of all the detail, one needs better understanding of the solar
model and the QM extension effect.

It is also interesting to note that if $\Delta \equiv m_1^2-m_2^2$ is very
small, for instance, which is close to the expected value for the vacuum
oscillation solution, i.e. the oscillation in medium cannot complete in one
cycle in the solar neutrino case, the effects of the extension term, no
matter from the micro black holes or the dense environment, increases the
transition probability from $\nu _e$ to $\nu _\mu$.

(ii) The interesting QM extension effect is possibly observable in other
neutrino oscillation experiments. The most likely possibility is in the
planned long baseline accelerator experiments, KEK-SuperKamiokande(250Km),
CERN-Gran Sasso(730Km) and Fermilab-Soudan2(730Km)\cite{Par}. The average
energies of $\mu $ neutrino beams are approximately 1 GeV, 6 GeV and 10 GeV
respectively. In all these long baseline experiments neutrino oscillation is
expected to be seen if the neutrino mass square difference is not much
smaller than $10^{-3}$ eV$^2$ and the mixing angle is not too small. With
the new QM extension effect being taken into account, we need to consider
two cases. If $\lambda $ is not much smaller than one, then the new effect
may modify neutrino oscillation significantly. And if $\lambda $ is much
larger than one, then the oscillation is observable for even smaller
neutrino mass square difference $10^{-3}\lambda ^{-1}$ eV$^2$. If the
distance of the experiment is chosen as $10^4$Km, that is about the diameter
of the Earth, the energy of the neutrino beam is about 1GeV, for neutrino
mass square difference as small as $10^{-5}$eV$^2$, the oscillation is
expected to be seen. So long as $\lambda $ is not too smaller than one, then
the QM extension effect is also to be observed for the mass squared
difference not too smaller than $10^{-5}$ eV$^2$.

(iii) Ellis et al. \cite{Ellis} and Reznik \cite{Reznik} introduced two
different mechanisms which both can induce transition (or oscillation) of
quantum systems from a pure state to a mixed one. Ellis et al. added the
extra $\delta H_{(4\times 4)}$ of $(4\times 4)$ form in the evolution
equation for $\rho^V$ of the four-vector form and this new term is not
hermitian. When one tries to turn the equation back to a $(2\times 2)$ form,
it is found that there is no way to make a Schr\"{o}dinger-type equation for
the $(2\times 2)\;\rho-$matrix such as $i({d/dt})\rho=H_{(2\times 2)}\rho$.
Reznik's formulation, instead, gives an equation for $\hat\rho$ of $2\times
2 $ form where he introduced the extension term $K\hat\rho K^{\prime}$ and
the equation also cannot be converted into an equation for $\rho$, even in
the four-vector form, but there $H_{(4\times 4)}$ is a hermitian matrix.
Generally the resultant $(2\times 2)$ $\hat\rho$ is not hermitian, while $%
\rho$ is. One can notice that the differences of Ellis et al.'s mechanism
from Reznik's explicitly.

Most generally, if $\rho$ (or $\hat\rho$) is in a four vector form, the
generalized Schr\"{o}dinger-type equation reads 
$$
i{\frac{d}{dt}}\rho^V=H_{(4\times 4)}\rho^V, \;\;\; ({\rm or\; for}\;
\hat\rho),%
$$
it can be converted into a $(2\times 2)$ matrix equation where $\rho$ (or $%
\hat\rho$) is a $2\times 2$ matrix as 
\begin{equation}
i{\frac{d}{dt}}\rho=L\rho+\rho R+L\rho K+\Delta\rho,\;\;\; (or\; for\;
\hat\rho), 
\end{equation}
where $L,R,K,\Delta$ are operators and can be decomposed as 
$$
O=O_0\sigma_0+O_i\sigma_i\;\;\;\; (O=L,\;R,\;K,\;\Delta;\;\; i=1,\;2,\;3)\;, 
$$
the coefficients $O_{\alpha} (\alpha=0,1,2,3)$ would be complex as the $%
L\rho(\hat\rho)K$ term exists. Writing out $H_{(4\times 4)}$, the real part
of $O_{\alpha}$ correspond to the hermitian components of $H_{(4\times 4)}$
while the imaginary parts to the anti-hermitian ones.

Because of the existence of the anti-hermitian components, the energy of the
system is not conserved as in Ellis et al.'s $H_{(4\times 4)}$. In our
formulation, the $H_{(4\times 4)}$ is hermitian, so the energy of the system
is conserved. Both the extra term $K\rho K^{\prime}$ and $K\hat\rho
K^{\prime}$ demand $(d/dt)Tr(\rho^2)\neq 0$, i.e. a system evolves from a
pure state into a mixed state.

(iv) We investigate the general properties of a quantum system which evolves
from a pure state into a mixed one due to either the micro black holes in
the background or certain environment in the case of neutrino oscillation.
With the generalized Schr\"{o}dinger equation, we have derived an expression
for $\hat\rho$ which is non-hermitian and the density matrix $%
\rho=\hat\rho\hat\rho^{\dagger}$. In this result, $\hat\rho\neq\hat\rho^{%
\dagger}$, naturally, one can decide that $\hat\rho^{\dagger}\hat\rho$
describes the environment \cite{Reznik}. In the case, $(d/dt)Tr\rho^2\neq 0$%
, thus the system evolves between pure and mixed states.

Thus we have derived the $\nu_e\rightarrow\nu_{\mu}$ transition rate and
find that the newly obtained expression is a modification of the famous
Mikheev-Smirnov-Wolfenstein's formulation.

In the modified expression, $\sin^2(2\theta_v) ({\frac{l_m}{l_v}})^2=
\sin^2(2\theta_m)$ keeps unchanged as in the MSW theory, but the oscillation
form $(1-\cos({\frac{2\pi t}{l_m}}))$ is modified into a more complicated
one as $(1-\cos({\frac{2\pi t}{l_m}})\cos({\frac{2\pi\lambda t}{l_m}}%
)+\cos(2\theta_v) \sin({\frac{2\pi t}{l_m}})\sin({\frac{2\pi\lambda t}{l_m}}%
))$. It obviously favors $\nu_e\rightarrow\nu_{\mu}$.

(v) The new extra term(s) either from micro black holes or from a matter
environment should be derived from certain theories in principle, whereas so
far no one has a reasonable way to derive it(them) out theoretically.
However as pointed out in the paper, the parameters, such as $\lambda$ for
instance, can be phenomenologically determined by fitting data if they fall
into a suitable region. The environmental influence may be observable in the
solar neutrino and/or the planned long-baseline experiments through neutrino
oscillations. Therefore we conclude that the environmental influence may be
experimentally accessible and certain constraints on the `extra' terms may
be set from the solar neutrino observation and the planned long-baseline
neutrino oscillation experiments in the foreseeable future.\\

{\bf Acknowledgement:} This work is supported in part by the National
Natural Science Foundation of China (NNSFC).


\vspace{2cm}

\begin{thebibliography}{99}
\bibitem{Smir}  A. Smirnov, Talk given at the International Workshop,
Particle Physics, Present and Future, Valencia, June, 1995, YITP-95-3.

\bibitem{Lan}  P. Langacker, ISSN 0418-9833.

\bibitem{Wol}  L. Wolfenstein, Phys.Rev. D17 (1978)2369.

\bibitem{Mik}  S. Mikheev and A. Smirnov, Sov.J.Nucl.Phys. 42 (1985)913.

\bibitem{Hir}  K. S. Hirata et al., Phys. Lett. 205B, 416 (1988); ibid 280B,
146(1992); D. Casper et al., Phys. Rev. Lett. 66, 2561(1991); R. Becker-
Szendy et al., Phys. Rev. D46, 3720(1992); NUSEX Collaboration, Europhys.
Lett. 8, 611(1989); ibid 15, 559 (1991); SOUDAN2 Collaboration, Nucl. Phys.
B35 (Proc. Suppl.); 427 (1994); ibid 38, 337 (1995); Fr\'ejus Collaboration,
Z. Phys. C66, 417 (1995); MACRO Collaboration, Phys. Lett. 357B, 481 (1995);
Y. Fukuda et al., Phys. Lett. 335B, 237 (1994).

\bibitem{Ellis}  J. Ellis, J. Hagelin, D. Nanopoulos and M. Sredwick,
Nucl.Phys. B241 (1984) 381.

\bibitem{Huet}  P. Huet and M. Peskin, Nucl.Phys. B434 (1995)3.

\bibitem{Hawking}  S.W. Hawking, Phys.Rev. D14 (1975)2460; Commun.Math.Phys.
87 (1982)395.

\bibitem{Reznik}  B. Reznik, Phys.Rev.Lett. 76 (1996)1192.

\bibitem{Vol}  M. Voloshin, M. Vysotskii and L. Okun, Sov.J.Nucl.Phys.44
(1986)440.

\bibitem{Smir1}  A. Smirnov, hep-ph/9611465; K. Benakli and A. Smirnov,
hep-ph/9703465.

\bibitem{Par}  S. Parke, Fermilab-Conf-93/056-T(hep-ph/9304271)(1993); L.
Camilleri, CERN preprint, CERN-PPE/94-87(1994); ICARUS Collaboration, Gran
Sasso Lab. preprint LNGS-94/99-I(1994); S. Wojcicki, invited talk at XVII
Conference on Neutrino Physics and Neutrino Astrophysics, June 13-19, 1996,
Helsinki.

\bibitem{Chang}  C.-H. Chang et al. in preparation.
\end{thebibliography}
\end{document}